\documentclass[%
reprint,
 amsmath,amssymb,
 aps,
 prl,
 showpacs,
 10pt,
 lengthcheck,
]{revtex4-1}
\usepackage{graphicx}
\usepackage{dcolumn}
\usepackage{bm}
\usepackage{hyperref}

\def\ket#1{\mathinner{|{#1}\rangle}}

\usepackage{siunitx}
\newcommand{\muin}{\mu_\text{in}}
\newcommand{\muout}{\mu_\text{out}}
\newcommand{\Ppol}{P_\text{pol}}
\newcommand{\Ptot}{P_\text{tot}}
\newcommand{\erbium}{$\text{Er}^{3+}$}
\begin{document}

\title{Quantum Cloning for Absolute Radiometry}
\author{Bruno Sanguinetti}
\email{Bruno.Sanguinetti@unige.ch}
\author{Enrico Pomarico}
\author{Pavel Sekatski}
\author{Hugo Zbinden}
\author{Nicolas Gisin}
\affiliation{Group of Applied Physics, University of Geneva, 20~rue de l'Ecole-de-M\'edecine, CH-1211 Geneva 4, Switzerland}


\begin{abstract}
In the quantum regime information can be copied with only a finite fidelity. This fidelity gradually increases to 1 as the system becomes classical. In this article we show how this fact can be used to directly measure the amount of radiated power. We demonstrate how these principles could be used to build a practical primary standard.
\end{abstract}
\maketitle

%
Since its inception quantum mechanics has had a deep tie with radiometry, the science of measurement of electromagnetic radiation. The electrical substitution radiometer, developed by Lummer and Kurlbaum in 1892~\cite{Lummer1892}, was used to observe the spectral distribution of a heated black body. In 1900 Max Planck was able to describe this distribution by assuming that electromagnetic radiation could only be emitted in multiples of an energy quantum  $E=h\,\nu$. This discovery not only provided an accurate law relating the radiated spectral density to temperature, but laid the foundations of quantum physics. The electrical substitution radiometer is still used as the primary standard for spectral radiance by many metrology laboratories. These systems have been improved over more than a century and can now achieve absolute uncertainties better than \SI{e-4}{}, when operated at relatively high powers~\cite{Houston06}.

More recently, nonlinear optical effects such as Spontaneous Parametric Down Conversion have provided a new primary standard based on the correlations of quantum fields~\cite{Poliakov2009}. The accuracy of these techniques has improved by nearly one order of magnitude every ten years, and is currently of the order of~\SI{e-3}{}. These systems are limited to the photon-counting regime, with a recent theoretical proposal for extension to higher photon rates~\cite{Brida09}.

In this letter, we present a radiometer that overcomes these limitations and works over a broad range of powers:  from the single photon level, up to several tens of \SI{}{nW} ($\approx \SI{e11}{}$ photons/s), i.e. from the quantum regime to the classical regime.
 In fact, our system is able to provide an absolute measure of spectral radiance by relying on a particular aspect of the quantum to classical transition: as the number of information carriers (photons) grows, so does the fidelity with which they can be cloned. For an optimal cloning machine~\cite{Simon00,DeMartini2000,DeMartini02,Fasel02,Lamas-Linares02} this relation can be derived \emph{ab initio}~\cite{Gisin97,Scarani05} so that a measurement of the fidelity of the cloning process is equivalent, as we shall see below, to an absolute measurement of spectral radiance.

Optimal cloning has been demonstrated in a variety of systems~\cite{DeMartini2000,Fasel02,Lamas-Linares02}. Stimulated emission in atomic systems is particularly practical as high gains can be easily achieved and the entire system can be implemented in-fibre which both ensures the presence of a single spatial mode and makes the system readily applicable, though not limited, to current telecom technology.

\emph{Principle of operation -- }
The aim of this experiment is to produce an absolute measurement of luminous power~$P_\text{in}$. We will do this by using an optimal Universal Quantum Cloning Machine (QCM). As we shall see such a device is able to directly relate a relative measurement of two orthogonal polarizations at its output to $P_\text{in}$. The relative measure that we use is the fidelity $\mathcal{F}$ which is the mean overlap between the input and output polarization, and can be expressed as follows:
\begin{equation}
\mathcal{F} = \frac{P_\parallel}{P_\parallel+P_\perp},
\end{equation}
where $P_\parallel$ and $P_\perp$ are the output powers in the polarizations parallel and perpendicular
to the polarization of the input light.

For an optimal QCM (losses are considered in the next section), the fidelity of a cloning process from $N$ to $M$ qubits can be derived \emph{ab initio} ~\cite{Gisin97} to be:
\begin{equation}
\mathcal{F}_{N\rightarrow M}=\frac{NM+N+M}{NM+2M}.
\end{equation}
This equation remains valid when we clone a large number of polarization qubits distributed over a large number of temporal modes and can be rewritten in terms of the average number of input and output photons per (temporal) mode $\muin$ and $\muout$~\cite{Fasel02}:
\begin{equation}
\mathcal{F}_{\muin\rightarrow\muout}
\equiv\frac{\mu_\parallel}{\mu_\parallel+\mu_\perp}
=\frac{\muin\muout+\muin+\muout}{\muin\muout+2\muout},
\label{eqn:fidelity_muin_muout}
\end{equation}
where $\muout$ contains both a number of exact copies of the input signal and intrinsic noise due to the amplification process, i.e. $\muout = \mu_\parallel+\mu_\perp$.

\begin{figure*}[t]
\begin{center}
\includegraphics[height=5cm]{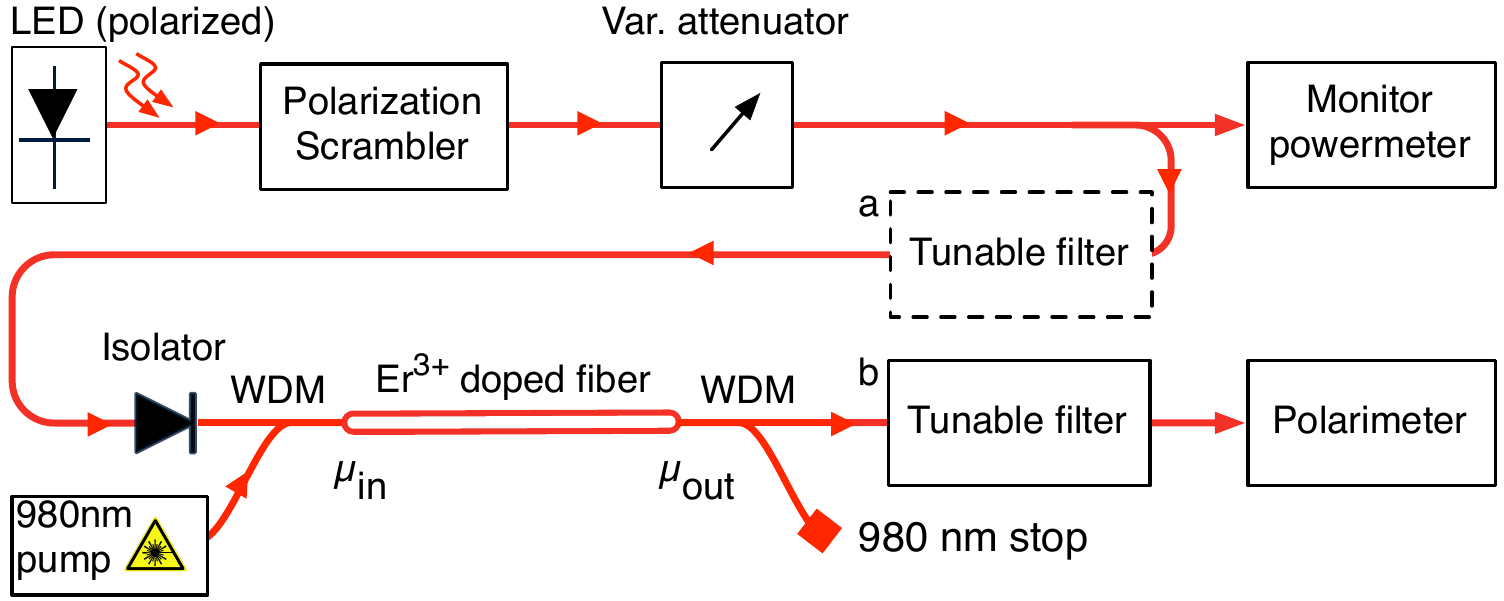}
\caption{Experimental arrangement: a broadband source with controllable polarization and power is amplified by an Erbium doped fiber amplifier. The degree of polarization (DOP) of the amplified light is then measured with a polarimeter. Spectral bandwidth is determined by a tunable filter. A value for the input power can be calculated from the DOP and compared with a calibrated powermeter (monitor).
}
\label{fig:setup}
\end{center}
\end{figure*}
It is also possible to express $\muout$ as a function of $\muin$ and the amplifier gain $G$~\cite{Shimoda57}:
$\muout$ is the sum of the stimulated emission $G\,\muin$ and the spontaneous emission, equivalent to amplifying the vacuum, so that:
\begin{equation}
\muout = G\,\muin+2(G-1).
\label{eqn:muout}
\end{equation}
Equations (\ref{eqn:fidelity_muin_muout}) and (\ref{eqn:muout}) can be combined to obtain the spectral radiance $\muin$ as a function of fidelity and gain:
\begin{equation}
\muin=\frac{2\,\mathcal{F}\,G-G-2\mathcal{F}+1}{G-\mathcal{F}\,G}\simeq \frac{2\mathcal{F}-1}{1-\mathcal{F}},
\label{eqn:fidelity}
\end{equation}
with the approximation holding for $G\gg 1$. $P_\text{in}$ can be derived from $\muin$ and a measurement of the number of modes per unit time $\tau_c^{-1}$.
%
%

Three aspects make this scheme attractive: the first is that after amplification input power information is polarization encoded and is therefore insensitive to losses
\footnote{The effect of Polarization Dependent Losses (PDL) is mitigated by averaging over a number of random polarizations produced by the scrambler.};
the second is that the experiment can be performed entirely in fiber, ensuring the selection of a single spatial mode. The third advantage is that this scheme works over a broad scale of powers: from single photon levels up to several tens of \SI{}{nW} ($\sim\SI{e11}{}$ photons/s).

\begin{figure}[htbp]
\begin{center}
\includegraphics[width=\columnwidth]{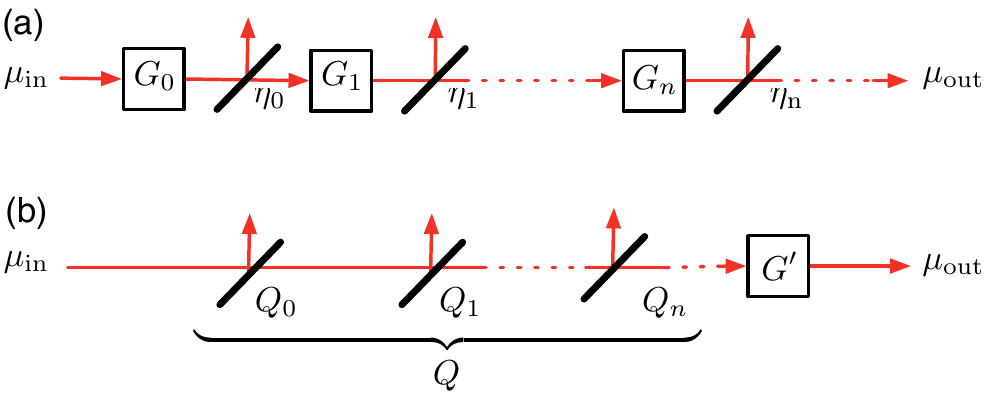}
\caption{(a) Model of a non-totally inverted medium as succession of infinitesimal gain elements $G_n$ and loss elements $\eta_n$. (b) Each loss element $\eta_n$ within the fiber is equivalent to a \emph{smaller} loss element $Q_n$ before the amplifier.}
\label{fig:model}
\end{center}
\end{figure}

\emph{Non-ideal cloning -- }
The reasoning presented above assumes the universal cloning process to be optimal. It has been shown theoretically that amplification in an inverted atomic medium indeed provides optimal cloning~\cite{Simon00}, but for precision applications it is important to consider the possible effects of a non perfectly inverted medium, which we model by a succession of infinitesimal gain and loss elements, $G_n$ and $\eta_n$, as shown in Fig.~\ref{fig:model}(a).
%
We have shown~\footnote{See supplementary material} that this model is equivalent to a much simpler one.
Each loss element $\eta_n$ can be represented by a different loss element $Q_n$ before an optimal cloning machine with gain $G'$, as shown in Fig.~\ref{fig:model}(b).
It can be shown that the product $Q$ of all $Q_n$ can be expressed as:
\begin{equation}
\label{eqn:Q}
Q=\prod Q_n\; ; \; Q_n=\frac{G_0^n\eta_n}{G_0^n\eta_n+(1-\eta_n)},
\end{equation}
where $G_0^n$ is the effective gain between the beginning of the amplifier and element $\eta_n$.
A fully inverted medium would have $Q=1$. %
From Eqn.~(\ref{eqn:Q}) it is apparent that the effect of a small loss ($\eta_n\lesssim 1$) is proportional to $1/G_0^n$. As $G_0^n$ grows exponentially over the length of the fibre, losses towards the end can be neglected. At the beginning of the amplifier two effects guarantee that the medium is fully inverted: the input signal is small, as it has not been amplified yet, and the signal and pump co-propagate, ensuring maximum pump power in this region. Cloning optimality can then be achieved in a non-ideal amplifier.
%

\begin{figure}[htbp]
\begin{center}
\includegraphics[width=\columnwidth]{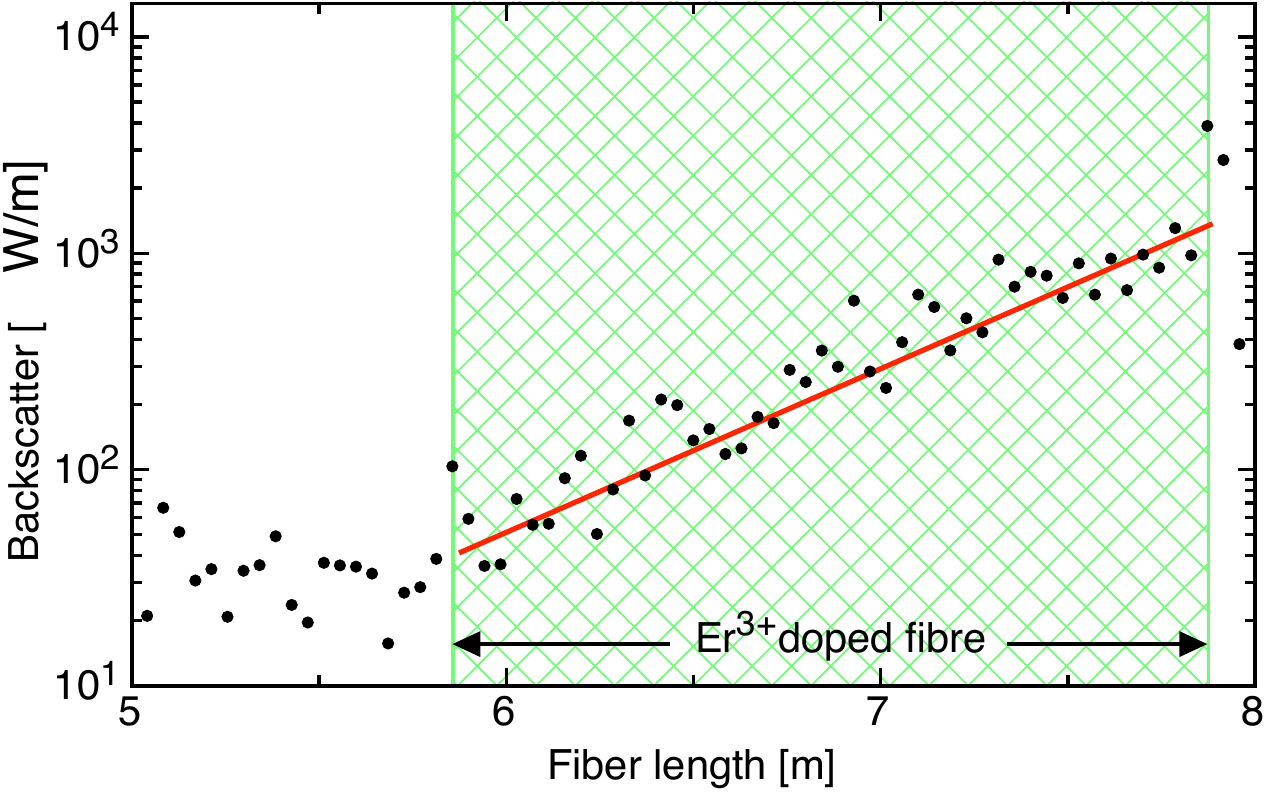}
\caption{Optical frequency-domain reflectometer measurement showing homogenous gain per unit length within the \erbium doped fibre. The solid line is an exponential fit of the data.}
\label{fig:ofdr}
\end{center}
\end{figure}

%
\emph{Experimental arrangement -- }
Fig. \ref{fig:setup} shows the setup, which can be conceptually divided in three main parts: generation of a set amount of power, amplification and fidelity measurement.
To test our system, we prepare states with a known number of photons per mode ($\muin$). This is done using a polarized LED that is passed through a polarization controller (scrambler) and a variable attenuator. The power is then split (50:50), with one branch monitored on a calibrated powermeter, while the other is sent to the amplification stage.
Amplification is provided by  \SI{2}{m} of $\text{Er}^+$ doped fiber, (attenuation \SI{16.7}{dB/m} at \SI{1530}{nm}), pumped by a \SI{980}{nm} diode laser. The pump light is combined with the signal on the input of the \erbium fibre using a wavelength division multiplexer (WDM), and an isolator is placed before the input to prevent unwanted resonances. After the \erbium doped fibre, most of the pump power is removed using an additional WDM. In this realization, the no-cloning theorem is guaranteed by the \erbium spontaneous emission, which adds randomly polarized photons to the signal. We used an optical frequency-domain reflectometer~\cite{Wegmuller:2000p798} to verify that the gain per unit length is constant over the entire fibre, indicating that the atomic medium is fully inverted. Results are shown in Fig.~\ref{fig:ofdr}.
The measurement stage  consists of a grating-based tunable filter and a polarimeter. The filter has a width of \SI{273.3(5)}{pm} (FWHM), which ensures that the effects of polarization mode dispersion can be neglected.
%
 %
%
The polarimeter measures the degree of polarization (DOP) with a nominal accuracy of 1\%, where the DOP is defined as the polarized power (in any basis) $\Ppol$ over the total power $\Ptot$, and is related to fidelity by $\mathcal{F}=(1+\text{DOP})/2$. Using a polarimeter rather than simply a polarizing beamsplitter and powermeter is less accurate, but allows us to test whether the system works equally well for arbitrary input states of polarization, i.e. whether the QCM is truly universal.

\emph{Experimental procedure -- }
To evaluate the accuracy of our system, we will need to compare our measurement of $\muin$ with the value $\muin^*$ obtained  from the reference powermeter. 
%
%
To do so, we first measure the ratio between the power at the monitor output and the power at the entrance of the amplifier within the bandwidth of the tunable filter. This is done by placing the filter just before the amplification stage (position `a' in Fig.~\ref{fig:setup}). Together with a measurement of the filter's attenuation and bandwidth, this allows us to obtain $\muin^*$ from   the monitor power. The filter is then placed after the amplification stage (position `b') so that \erbium spontaneous emission outside of the bandwidth of interest is eliminated.
We then vary $\muin^*$ using the attenuator, and record the monitor power versus the degree of polarization. 
For each $\muin^*$ the measurements are repeated for 20 different input polarizations, to estimate uncertainties.

\begin{figure}[htbp]
\begin{center}
\includegraphics[width=\columnwidth]{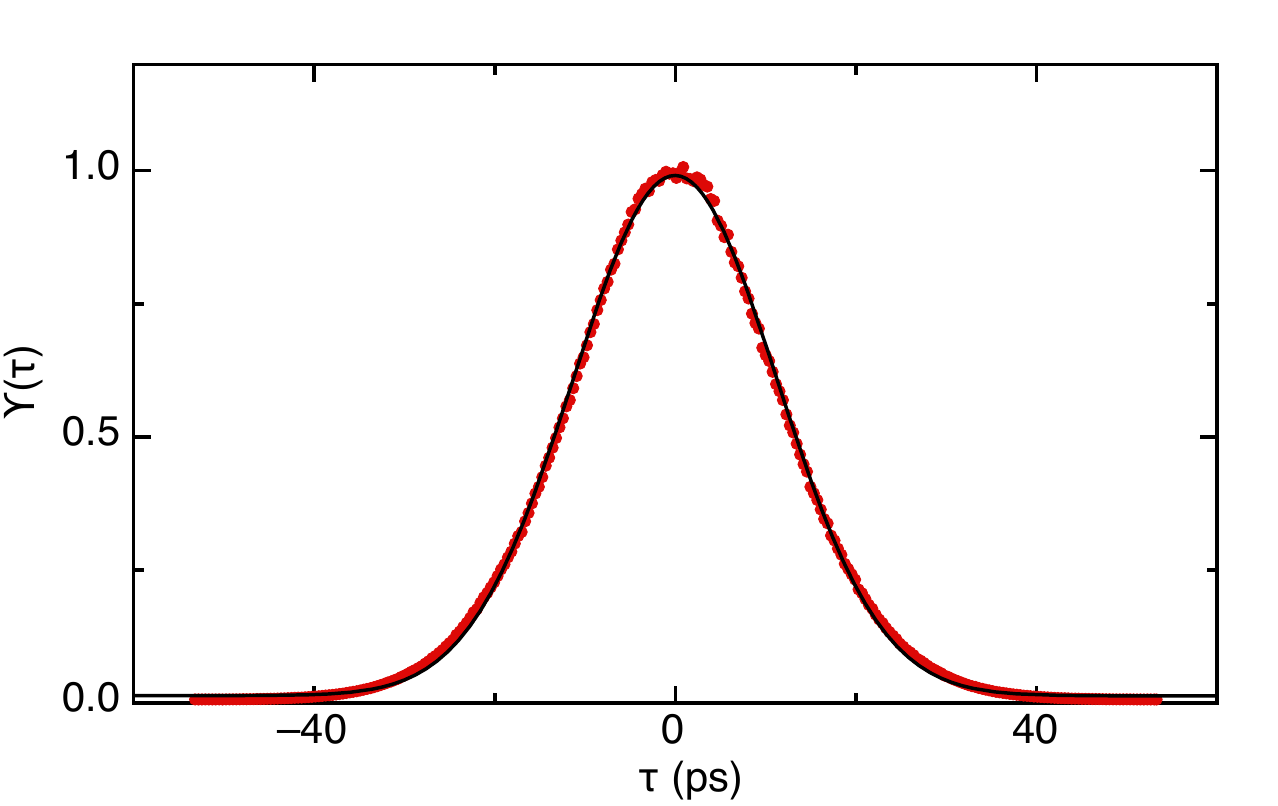}
\caption{Autocorrelation function of the source after the filter. $\gamma(\tau)$ is the fringe visibility measured with a low-coherence interferometer; $\tau_c$ will simply be the numerical integral of this data. The solid line is a Gaussian fit.}
\label{fig:autocorrelation}
\end{center}
\end{figure}

\emph{Results --}The fidelity of the cloning process is a measure of spectral radiance. In order to measure power it is necessary to have an accurate measure of the number of modes involved. Using a single-mode fibre ensures that there is only a single spatial mode: only the number of temporal modes per second need to be measured. It is convenient to define the coherence time $\tau_c$  as in~\cite{Mandel62}:
\begin{equation}
\tau_c = \int_{-\infty}^{\infty}|\gamma(\tau)|^2\,d\tau
\end{equation}
where $\gamma(\tau)$ is the autocorrelation function normalized such that $\gamma (0)=1$. Using this definition, the coherence length $c\,\tau_c$ is the length of the unit cell of photon phase space~\cite{Mandel62}, so that the number of modes per second is simply~$\tau_c^{-1}$. Measuring this value with an optical low-coherence interferometer (Fig.~\ref{fig:autocorrelation}) yields $\tau_c=\SI{19.71(4)}{ps}$ which corresponds, assuming a Gaussian shape, to wavelength FWHM of $\Delta\lambda=\SI{273.3(5)}{pm}$. We also performed a (less precise) spectrometric measurement yielding $\Delta\lambda=\SI{271}{pm}$. With this filter, a mean of one photon per temporal mode corresponds to \SI{6.461}{nW}.
%
%
%
\begin{figure}[htbp]
\begin{center}
\includegraphics[width=\columnwidth]{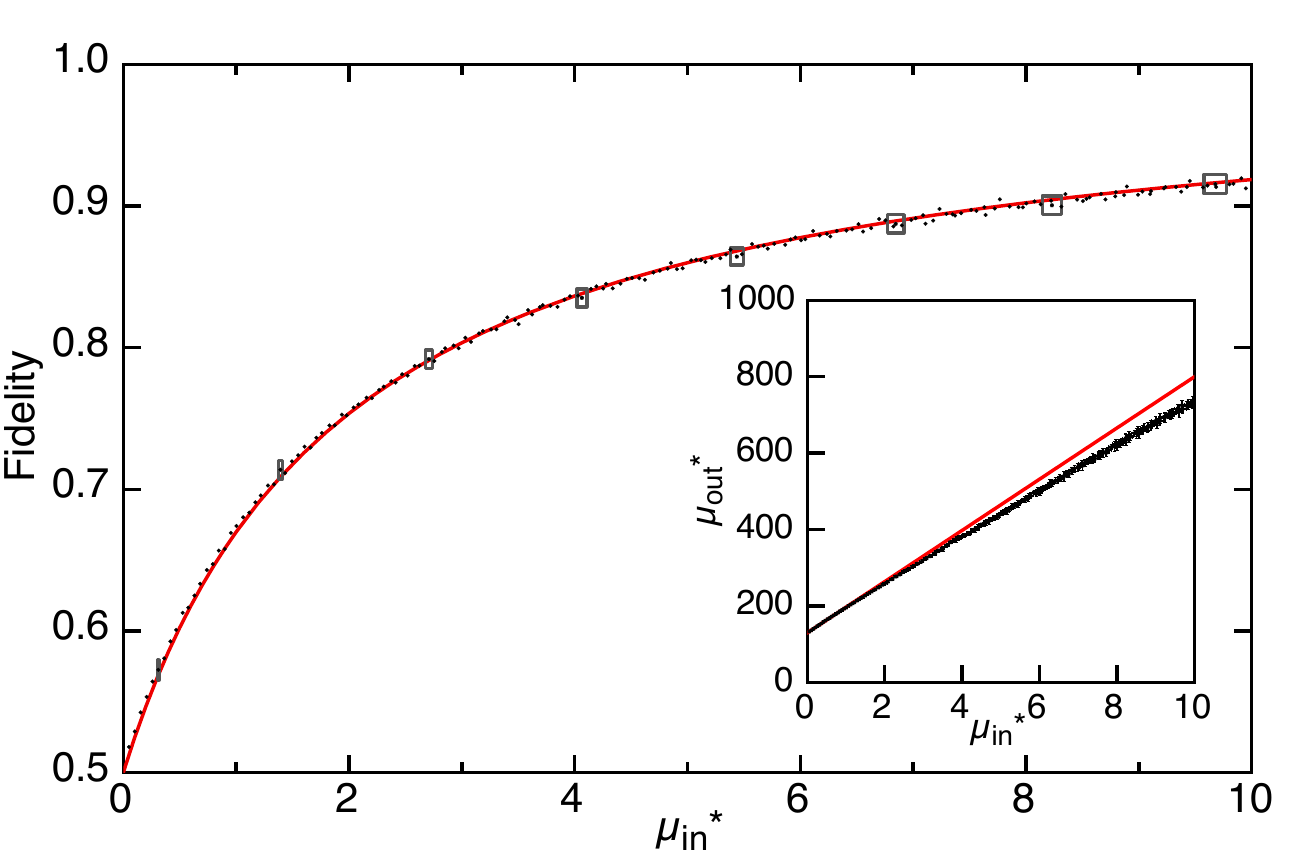}
\caption{Fidelity versus number of input photons per mode, fitted with Eqn.~(\ref{eqn:fidelity}). Representative errors are shown as boxes on some points. The inset shows the output versus input number of photons per mode, the line is a fit on the first data points ($\muin^* < 1$), showing reduced gain as $\muin$ grows. }
\label{fig:combined}
\end{center}
\end{figure}

We measure the amplifier gain by directly comparing the power at the output of the amplifier with the power at the input. The inset of Fig.~\ref{fig:combined} shows a typical plot, in terms of $\muin^*$ and $\muout^*$,  the thickness of the line represents random errors.  Note that the gain is $G=\partial\muout / \partial\muin$; so that any systematic error in either the power measurement or the estimation of the number of modes cancels. The line in the inset of Fig.~\ref{fig:combined} is a fit of the data for $\muin^*<1$, revealing that at high $\muin$ the gain is reduced. This effect could be minimized by pumping from both sides of the \erbium doped fiber. Nevertheless, the gain is constant for $\muin<2$, allowing us to assume within this range that the intercept $\mu_0$ corresponds to the spontaneous emission ($2\,G-2$) from Eqn.~(\ref{eqn:muout}), so that: $\muout =   G\,\muin + \mu_0$.
In this range it is then possible to measure $\muin$ without distinguishing the polarizations, as $\muin = 2(\muout^*-\muin^*)/\mu_0^* -2 $.

We then measure the fidelity $\mathcal{F}$ versus  $\muin^*$; Fig.~\ref{fig:combined} shows a typical plot, which can be fitted with Eqn.~(\ref{eqn:fidelity}), where $\muin$ has been replaced with $k\muin^*$, and $k$ is the fitted parameter. With this definition $k = \muin/\muin^*$ represents the discrepancy between our measurement of $\muin$ and the value $\muin^*$ obtained from the reference powermeter. Here, $k$ also accounts for the possibility of non-optimal cloning which would introduce a further factor $Q\le1$ equivalent to a loss on the input of the cloning machine.
The fitted curve in Fig.~\ref{fig:combined} yields $k=\SI{1.013(5)}{}$, where the error indicated represents statistical uncertainty. Systematic errors, as we shall see in the next section, could be up to one order of magnitude higher.

\emph{Error estimation -- }
The aim of this experiment was to demonstrate the principle of a cloning radiometer, rather that to build a standard that can compete with metrology laboratories. It is however important to discuss the errors involved, both for the interpretation of the results and to evaluate the applicability of this method.
One of the advantages of this technique is that relative measurements, which usually have small errors, are used; but how does a small uncertainty in the fidelity $\Delta\mathcal{F}$ translate into an error in the measurement of $\muin$? From  Eqn.~(\ref{eqn:fidelity}) we obtain:
\begin{equation}
\Delta\muin=(2+\muin)^2\Delta\mathcal{F},
\label{eqn:errors}
\end{equation}
so that  $\Delta\muin/\muin$ has a minimum of $\Delta\muin/\muin=8\Delta\mathcal{F}$ at $\muin=2$, i.e. when spontaneous and stimulated emissions are equal. At higher spectral radiances, $\Delta\muin/\muin$ rises linearly with $\muin$. The spectral bandwidth of the filter can be chosen as to operate in the desired power regime: our system is optimal at \SI{13}{nW}, commercially available filters would allow this point to be easily lowered to \SI{100}{pW}. From preliminary tests we estimate that this technique would work to an upper limit of \SI{100}{nW}, above which the effects of polarization mode dispersion and wavelength dependence of the components need to be taken into account.

The two main systematic uncertainties in our system are due to the reference powermeter, and to the polarization measurement.
The powermeter is an EXFO PM-1100, recently calibrated by METAS to an absolute uncertainty of 0.7\% and with a measurement to measurement standard deviation of 0.5\% (including fibre re-connection). The linearity of this powermeter is within this uncertainty over its entire range. The fidelity is measured with a Profile PAT~9000 polarimeter which has a nominal $\Delta\mathcal{F}=0.5\%$. We noticed that the fidelity was overestimated by 1\% for unpolarized light, and underestimated by 0.2\% for polarized light.
Systematic errors are dominated by the polarimeter, so that $\Delta\muin/\muin\sim 4\%$ for $\muin=2$.

\emph{Conclusion -- }
We have shown that the fidelity of cloning can be used to produce an absolute power measurement  with an uncertainty only limited by the uncertainty of a relative power measurement. We demonstrate the scheme with an all-fiber experiment at telecom wavelengths, achieving an accuracy of 4\%, with much space for improvement by a metrology laboratory. The experiment also demonstrates optimal $1\rightarrow 67$ cloning, and is an interesting application of Quantum Information Science, and in particular of the study of the quantum to classical transition.

\emph{Acknowledgements}
We are very grateful to Jacques Morel and  Armin Gambon of the Swiss Federal Office of Metrology (METAS) for useful discussion and for the calibration of our reference powermeter.  As always, we thank Claudio Barreiro and Olivier Guinnard for their technical insights. Financial support for this project was provided by the Swiss NCCR-QP and by the European Q-ESSENCE project.


\bibliography{papers} 

\appendix

\section{Appendix A: treatment  of potential losses in a doped fiber amplifier}
\begin{figure}[h]
  \center
  \includegraphics[width=8cm]{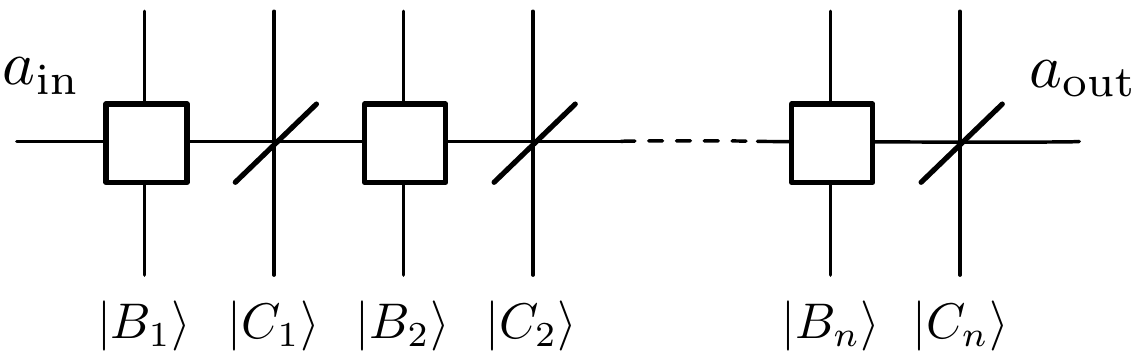}\\
  \caption{Symbolical representation of the lossy amplification of mode $a$ as a sequence of distinct amplifying elements (rectangles) spaced out by beam splitters ($\textbf{/}$ signs). $\ket{B_n}$ and $\ket{C_n}$ stand for the initial states of the local auxiliary modes.}\label{fig:amplification_model}
\end{figure}

In order to evaluate the feasibility of optimal universal quantum cloning via stimulated emission in an Er$^{3+}$ doped fiber, we should take into account the potential effects of internal losses.
The amplification medium can be naively modeled, as shown in Fig.~\ref{fig:amplification_model}: a sequence of thin amplifying
atomic slices spaced out by beam splitters, representing the internal optical losses. The propagation of the photonic mode in a lossy amplifier can be seen as successive interaction with these elements.

The interaction of the input propagation mode $a$ with the $n$-th beam-splitter can be represented by the hamiltonian $H^L_n=i\lambda(a c_n^{\dag}-a^{\dag}c_n)$, where `$L$' stands for losses, $\lambda$ is a constant and $c_n$ is an auxiliary mode initially in the vacuum state ($\ket{C_n}= \ket{0}_{c_n}$). By using the time evolution operator $U_n^L=e^{-\frac{iH_{n}^{L} t}{\hbar}}$, in the Heisenberg picture the action of the $n$-th beam-splitter on modes $a$ and $c_n$ can be expressed by the relation
\begin{eqnarray}\label{eq:loss_element}
a^{L} = U_n^{L\dag}a U_n^L=\sqrt{\eta_n}\, a + \sqrt{1-\eta_n}\, c_n,
\end{eqnarray}
where $\eta_n$ is the specific transmission coefficient of the beam-splitter element.

A similar relation can be found for the amplifying element.
It has been shown in \cite{Kempe:2000p10} that amplification in an inverted atomic medium provides optimal universal cloning, equivalent to stimulated parametric down conversion in nonlinear crystals. For this reason the interaction of the propagation mode $a$ with an amplifying atomic element can be expressed more conveniently by the hamiltonian of the parametric amplification process, which for the $n$-th amplifying element is $H_n^A = i\chi (a\, b_n - a^\dag b_n^\dag)$, where $\chi$ is a constant and `$A$' stands for `amplification'. $b_n$ in the parametric case is the mode of the anticlones \cite{Simon00}, while in the atomic case it represents a collective ``desexcitation'' of the atoms in the $n$th-slice \cite{Kempe:2000p10}. Initially $\ket{B_n}=\ket{0}_{b_n}$, meaning that the population is inverted in the element (all the atoms are in the excited state). The amplifying elements are characterized by different gain values $G_n$. Notice that the hamiltonian $H_n^A$ has a second term containing $a_\perp$ that ensures universality, however this term is decoupled and doesn't affect mode $a$.

\begin{figure}[h]
\center
  \includegraphics[width=7 cm]{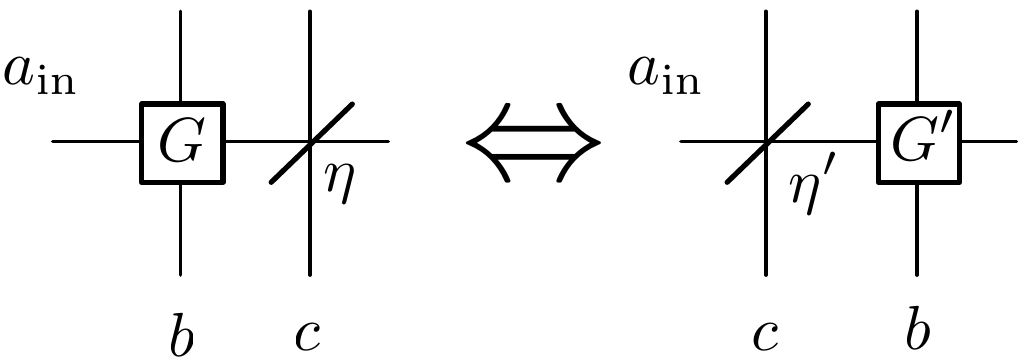}\\
  \caption{The situations where the loss comes before or after the amplification are physically equivalent if the scalar parameters $(G,\eta)$ and $(G',  \eta')$ satisfy (\ref{trololo})}
  \label{fig:equivalence}
\end{figure}

By using the time evolution operator $U_n^A=e^{\frac{-iH_n^A t}{\hbar}}$, in the Heisenberg picture the action of the amplifying element on modes $a$ and $b_n$ yields:
\begin{eqnarray}\label{eq:amplifying_element}
a^{A} = U_n^{A \dag}a U_n^A = \sqrt{G_n}\, a + \sqrt{(G_n -1)}\, b_n^\dag.
\end{eqnarray}
Now let us consider the two different situations represented schematically in Fig.~\ref{fig:equivalence}. In the first case the propagation mode $a_{in}$ is through an amplifying element of gain $G$ before interacting with mode $c$ in a beam splitter of transmission $\eta$. In the second situation the order is inverted with the parameters $G'$ and $\eta'$. Using (\ref{eq:loss_element}) and (\ref{eq:amplifying_element}) the value of $a$ at the output for the two cases is:
\begin{align}
a_{out}^1 &=\sqrt{G \eta}\, a_{in} + \sqrt{\eta (G-1)}\, b^\dag + \sqrt{1-\eta}\, c,
\\
a_{out}^2 &=\sqrt{G' \eta'}\, a_{in} + \sqrt{(G'-1)}\, b^\dag + \sqrt{G'(1-\eta')}\, c.
\end{align}
Suppose that we fix the value of $G$ and $\eta$ and solve for the value of $G'$ and $\eta'$ that would give the same output $a_{out}^1 = a_{out}^2$. The following three conditions must be satisfied
\begin{equation}\label{eq:conditions}
\left\{
    \begin{array}{ccc}
    G \eta &=& G'\eta',\\
    \eta (G-1) &=& (G'-1),\\
    1-\eta &=& G'(1-\eta').
\end{array}
\right.
\end{equation}

Since the first equation in (\ref{eq:conditions}) is the difference of the other two, the system always has
the solution
\begin{equation}\label{trololo}
\left\{
    \begin{array}{ccc}
    G' &=& \eta(G-1)+1,\\
    \eta'&=& \frac{G\eta}{G\eta +(1-\eta)}.\\
\end{array}
\right.
\end{equation}

\begin{figure}[h]
\center
  \includegraphics[width=8 cm]{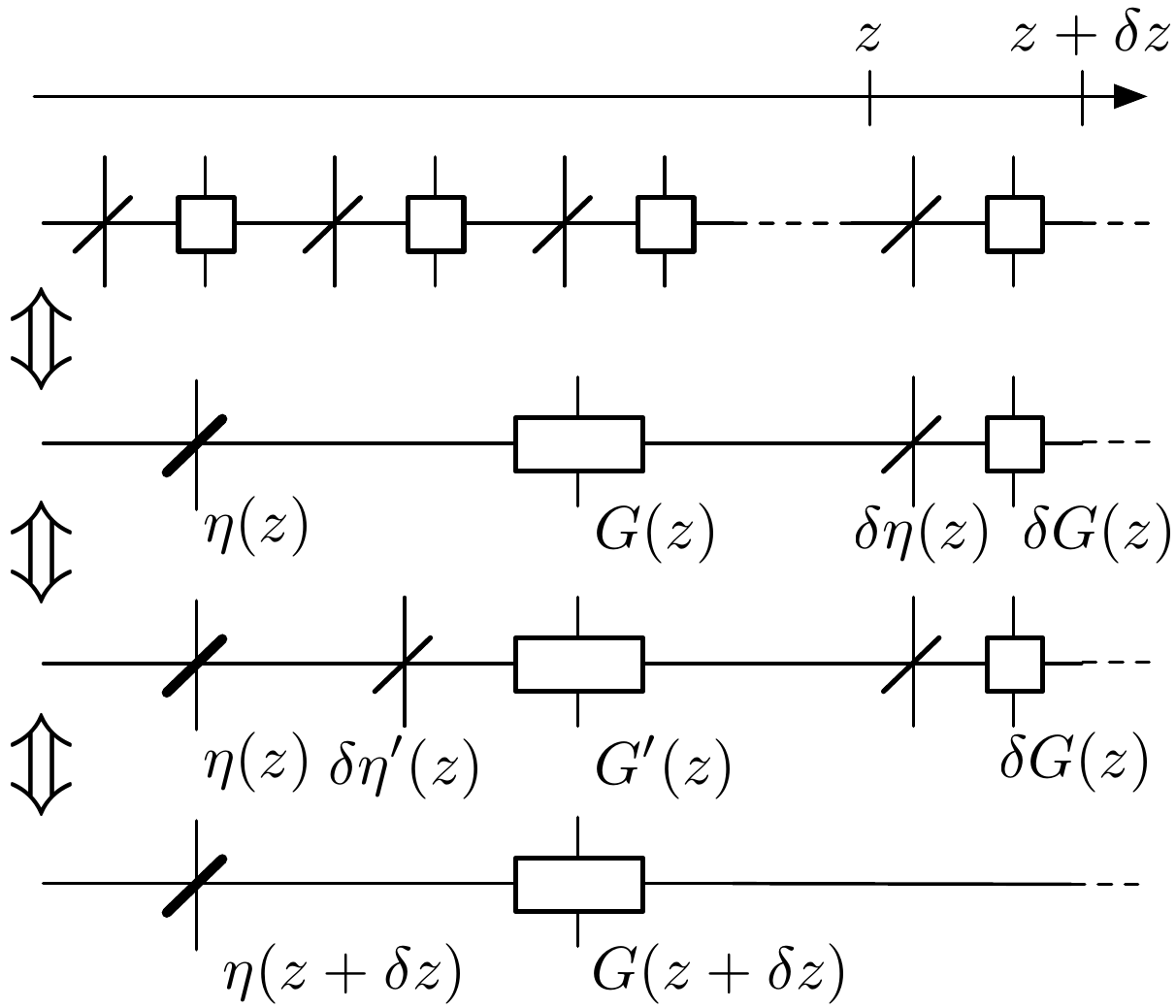}\\
  \caption{The infinitesimal loss and gain elements are successively absorbed in a global gain parameter $G(z)$ and a global loss parameter $\eta(z)$. In the limit $\delta z \rightarrow 0$ Eqns. (\ref{trololo}) yield differential equations for these functions.}
 \label{cont}
\end{figure}

It is easy to verify that this solution satisfies $G'\geq 1$ and $0 \leq  \eta' \leq 1$ for any given $\eta$ and $G$, while the inverse is not true. The last equation in (\ref{eq:conditions}) clearly stipulates that the condition $G'(1-\eta')< 1$ must be satisfied
if we want to rearrange the elements (else it would imply a negative transmission $\eta$).
The consequence of this result is that we can pull all the beam-splitter elements in Fig.~1 on the left if we take care of correctly modifying the characteristic parameter for each element. It is well known that a combination of beam-splitter interactions with modes $c_i$ is equivalent to an interaction with a single mode $\tilde{c}$ being a linear combination of $c_i$ with the resulting transmission rate $\tilde{\eta} = \prod_i \eta_i$, the same is valid for a series of amplification layers implying $\tilde{b}$ and $\tilde{G} = \prod G_i$. So the initial process represented in Fig.~1 can be equivalently seen as a transmission loss $\tilde{\eta}'$ before an amplification $\tilde{G}'$, or the other way around with $\tilde{G}$ and $\tilde{\eta}$ if $\tilde{G}'(1-\tilde{\eta}')< 1$ is satisfied. Another point is that the lossy elements at the beginning of the propagation line will contribute more to the resulting effective loss that those at the end, naively one can see from (\ref{trololo}) that pulling a small loss element $\eta_n  = 1-\varepsilon$ on the left through $G_1,\,G_2,\,\ldots,\,G_n$ results in an effective loss of:
\begin{equation}
\eta'_n = \frac{G_1\ldots G_n \eta_n}{G_1\ldots G_n \eta_n+ (1-\eta_n)}\approx 1- \frac{\varepsilon}{G_1\ldots G_n},
\end{equation}
which illustrates the fact that in real experiments it is much better to have a copropagating pump laser.
 To be more quantitative
we can introduce continuous fields $\chi(z)$ and $\lambda(z)$, in such a way that propagating from $z$ to $z+\delta z$ the mode $a$ undergoes an infinitesimal gain $\delta G(z) = 1+\chi(z)\delta z$ and an infinitesimal transmission loss $\delta \eta(z) =1-\lambda(z)\delta z$ (for two infinitesimal elements the order is not important) as illustrated in the first line of the Fig.~\ref{cont}. Then starting from the left we put all the loss elements before the gain elements. We then get a resulting equivalent loss at point $z$ of $\eta(z)$ and a resulting equivalent gain $G(z)$, that would be the second line of the Fig.~\ref{cont}.
Then using the equations (\ref{trololo}) we permutate $G(z)$ with $\delta \eta(z)$ obtaining $G'(z)$ and $\delta \eta'(z)$; and finally combine them into $\eta(z+\delta z)= \eta (z) \delta \eta'(z)$ and  $G(z+\delta z) = G'(z)\delta G(z)$ (last line in Fig.~\ref{cont}) to obtain:
\begin{equation}
\left\{
    \begin{array}{ccc}
    G(z+\delta z) &=& \eta(z)(\frac{G(z)\delta\eta(z)}{G(z)\delta\eta(z)+ 1-\delta\eta(z)} ) \\
    \eta(z+\delta z) &=& (\delta\eta(z) (G(z)-1)+ 1)\delta G(z)
\end{array}
\right.
\end{equation}
In the limit $\delta z \rightarrow 0$ expanding the expressions in the r.h.s. to the first order in $\delta z$ we obtain the system of differential equation that $G(z)$ and $\eta(z)$ obey.
\begin{equation}\label{diff}
\left\{
    \begin{array}{ccc}
    \partial_z G(z) &=& (\chi(z)-\lambda(z))G(z)+\lambda(z)\\
    \partial_z \eta(z) &=& -\eta(z) \frac{\lambda(z)}{G(z)}
\end{array}
\right.
\end{equation}
the solution of this system is:
\begin{equation}
\left\{
    \begin{array}{ccc}
     G(z) = e^{\int_0^z(\chi(z_1)-\lambda(z_1))dz_1}[\int_0^z \lambda(z_1)e^{-\int_0^{z_1}(\chi(z_2)-\lambda(z_2))dz_2}+1]\\
     \eta(z) = (1+\int_0^z \lambda(z_1)e^{-\int_0^{z_1}(\chi(z_2)-\lambda(z_2))dz_2})^{-1}
\end{array}
\right.
\nonumber
\end{equation}
This solution can be used to take into account the effect of any arbitrary loss profile. For our current system the effect of losses is negligible, however it is important to know that for future  systems the loss characteristics of the fiber and amplification process can be fully taken into account by measuring only two parameters of the fiber: $G$ and $\eta$, and will thus not impose a limit on the accuracy of a high precision system.

To illustrate the solution of this equation we consider a sample with three different profiles of atomic inversion, namely $\chi_1(z)=2-e^{+\alpha z}$, $\chi_2= cte$ and $\chi_3(z)= 2- e^{\alpha(L-z)}$. 
The total equivalent loss $\eta$ for each of these three profiles is given in Fig.~4.

\begin{figure}[h]

\center
  \includegraphics[width=\columnwidth]{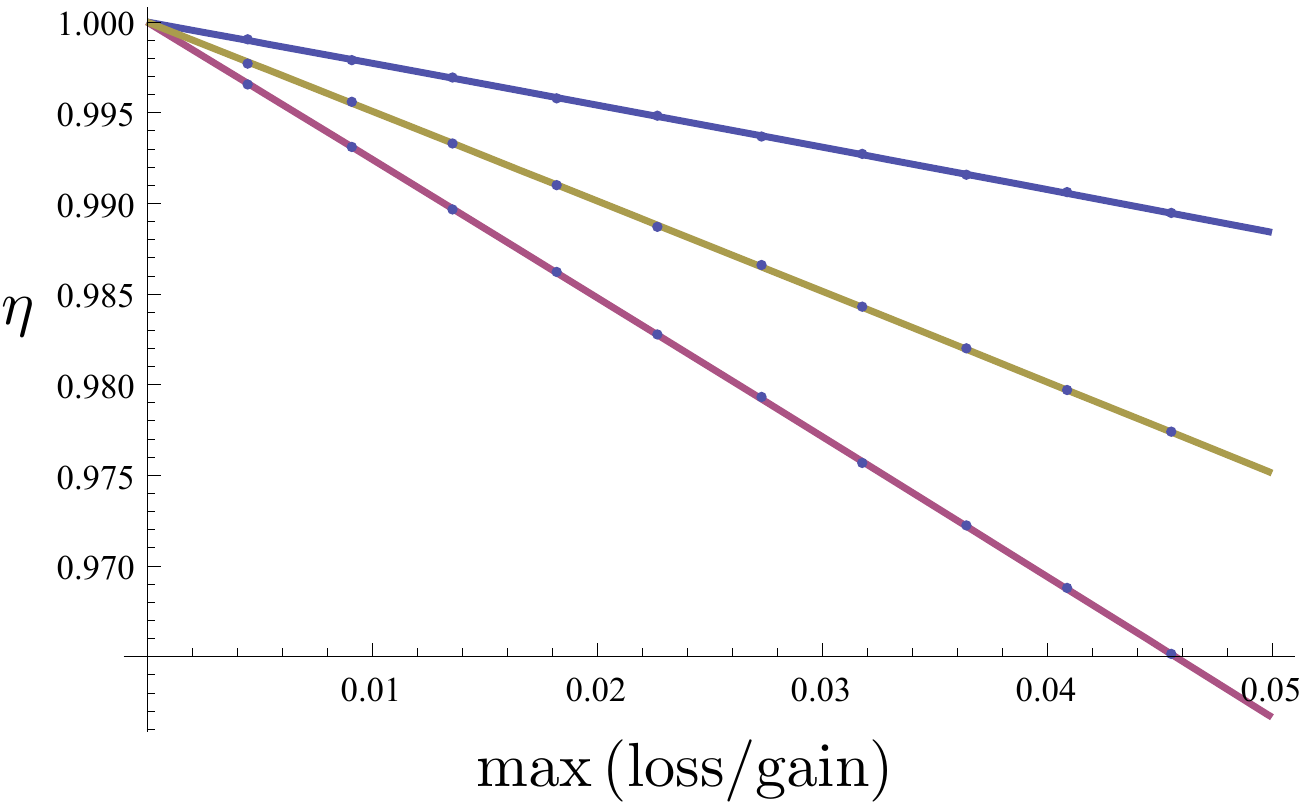}\\
  \caption{Global resulting loss $\eta$ for, from top to bottom, $\chi_1$, $\chi_2$ and $\chi_3$ as a function of $\alpha$. The total equivalent gain for these plots is approximately 50.}
\end{figure}

\end{document}